\documentclass[intlimits,twoside,a4paper]{article}
\usepackage{amsmath,amssymb}
\usepackage{graphicx}
\usepackage{wrapfig}
\usepackage[T2A]{fontenc}
\usepackage[cp1251]{inputenc}
\usepackage{color}

\usepackage{cmpj2}

\issue{2013}{16}{3}{33001}
\doinumber{10.5488/CMP.16.33001}

\title[Properties of spatial arrangement of V-type defects]%
{Properties of spatial arrangement of V-type defects in irradiated materials: 3D-modelling}%

\author[V.O. Kharchenko, D.O. Kharchenko]{V.O. Kharchenko, D.O. Kharchenko}
\address{Institute of Applied Physics of the National Academy of Scienses of Ukraine,\\
58 Petropavlivska St., 40000 Sumy, Ukraine}

\date{Received March 11, 2013, in final form May 20, 2013}
\authorcopyright{V.O. Kharchenko, D.O. Kharchenko, 2013}

\begin{document}

\maketitle

\begin{abstract}
We consider the dynamics of pattern formation in a system of point defects under
sustained irradiation within the framework of the rate theory. In our study we
generalize the standard approach taking into account a production of defects by
elastic fields and a stochastic production representing internal
multiplicative noise. Using 3D-modelling we have shown that with the damage rate
growth, a morphology of clusters composed of vacancies changes. The same
effect is observed with variation in the multiplicative noise intensity.
Stationary patterns are studied by means of correlation analysis.
\pacs 05.40.-a, 61.72.-y, 81.16.Rf

\end{abstract}

\vspace{2mm}
\section{Introduction}

It is well known that metals and alloys under irradiation are typical examples
of far-from-equilibrium systems with nonlinear coupling between their elements.
The simplest ones are vacancies (V-defects) and interstitials (I-defects).
Depending on irradiation conditions related to both defects replacement rate
and temperature, the point defects can arrange into defects of higher dimension
such as clusters, defect walls with vacancy loops \cite{118,1110}, voids
\cite{111}, precipitates \cite{117}, bubble lattices \cite{114,115,116}. At the
same time, irradiation can modify a surface of targets due to sputtering
resulting in the formation of ordered nano-size surface structures widely used in
modern electronics \cite{PRE2010,CMP2011,PRE2012}. Such patterns are considered
to be dissipative structures. Their theoretical and experimental investigations
are widely discussed in literature during last four decades (see, for example,
\cite{Martin,116,1112,Ryazanov,AbromeitMartin1999,Sugakov95}). The production of
defects and their microstructure can principally modify the physical and mechanical
properties of an irradiated material leading to swelling, hardening, etc. As
far as defect clusters are responsible for material embrittlement
\cite{Osetsky,Gavini}, a study of microstructure of materials, defects and their
arrangement into clusters is an urgent problem in modern material science and
in statistical physics. In the present paper we consider the processes of the formation of V-type
defect structures using both analytical treatment and numerical
modelling.

In theoretical investigations of microstructure formation, a formalism known as
multiscale modelling serves as an effective tool. In the framework of this
approach, properties of the system on lower (microscopic) hierarchical level are
self-consistently taken into account in the modelling procedures at higher levels of
description (mesoscopic or macroscopic). On the other hand, one can use the
so-called ``hybrid'' methods such as phase field crystals method where the dynamics
of microscopic crystal system is described in terms of atomic densities covering
time scales for elastic interactions between atoms and their diffusion
\cite{ElderGrant2004,Tupper}. It can be used to study the diffusion of defects
\cite{Elder2006}, dislocation dynamics \cite{EPBSG2007,UJP2012} and structural
transitions \cite{BEG2008,JAEA2009}. At a mesoscopic hierarchical level, one can
consider the dynamics of defects using the rate theory where concentration of both
interstitials and vacancies are considered (see, for example,
references \cite{Martin,AbromeitMartin1999,1112,EPJB2012}). Usually such models take
into account irradiation production rate and interactions between defects
except defect production caused by elastic deformation of a medium. An idea
of stochastic production of defects during irradiation was exploited
to study the ordering, chemical patterning and phase separation of irradiated
materials
\cite{Abromeit96,Enrique63,YeBellon04_1,EPJB2010,PhysA2010_1,PhysA2010_2,CEJP2011}.

The main goal of our paper is to consider the dynamics and spatial arrangement of
an ensemble of point defects using the rate theory describing the evolution of point
defects. We take into account the production of defects by irradiation, relaxation
(the effect of sinks) and the production of defects by elastic deformation of a medium
caused by point defects. We generalize the standard approach \cite{UFN1996}
considering a system with two spatial scales related to the diffusion and
microscopic interactions. Within the framework of the previously derived approach used
to study nanosize patterns in stochastic systems (see
references \cite{Haeften2004,MHW2005,PhysicaD,Mangioni2010}) we consider the most
probable stationary defect structures. In numerical investigations we solve
a three-dimensional problem and discuss the effect of the system parameters onto
the dynamics and morphology change of defect structures. We describe stationary
patterns by means of correlation analysis.

The paper is organized as follows. A generalized model described by the main
mechanisms of the formation of defects with two spatial scales related to diffusion
scale and defect interaction scale is presented in section~2. The analysis of
stationary states and their stability is provided in section~3. In section~4
we numerically study the pattern formation where we discuss the dynamics of statistical
averages and analyze stationary patterns. We conclude in section~5.

\section{Model and basic equations}

Following the standard approach \cite{Walgraef} the dynamics of point defects
is described by a two-component model
\begin{align}\label{cvci}
%\begin{split}
 \partial_{\mathrm{t}} c_{\mathrm{v}}&=K-D_{\mathrm{v}}S_{\mathrm{v}}(c_{\mathrm{v}}-c_{\mathrm{v}}^0)-\alpha c_{\mathrm{v}}c_{\mathrm{i}}\, ,\nonumber \\
 \partial_{\mathrm{t}} c_{\mathrm{i}}&=K-D_{\mathrm{i}}S_{\mathrm{i}}c_{\mathrm{i}}-\alpha c_{\mathrm{v}}c_{\mathrm{i}}\, .
%\end{split}
\end{align}
Here, $c_{\mathrm{v}}$ corresponds to populations of vacancies ($c_{\mathrm{v}}^0$ is the equilibrium
vacancy concentration) and $c_{\mathrm{i}}$ relates to interstitials. The first terms in
equation (\ref{cvci}) relate to the displacement damage rate and take into account a
production of defects due to irradiation. The second terms describe
the effect of sinks ($S_{\mathrm{i}}$ and $S_{\mathrm{v}}$) related to bias factors $Z_{\mathrm{i},\mathrm{v}}$, network
dislocation density $\rho_N$, vacancy loops $\rho_{\mathrm{v}}$, and interstitial loop
densities $\rho_{\mathrm{i}}$ as follows: $S_{\{\mathrm{v,i}\}}=Z_{{\{\mathrm{v,i}\}} N}\rho_{N}+Z_{{\{\mathrm{v,i}\}}
V}\rho_{\mathrm{v}}+Z_{{\{\mathrm{v,i}\}} I}\rho_{\mathrm{i}}$, where $Z_{\mathrm{v}N}=Z_{\mathrm{v}I}=Z_{\mathrm{v}V}=1$,
$Z_{\mathrm{i}N}=1+B$, $Z_{\mathrm{i}I}\simeq Z_{\mathrm{i}V}\simeq1+B'$, $B'\geqslant B$, $B\simeq 0.1$.
Diffusivities of vacancies and interstitials are defined in the standard
manner: $D_{\{\mathrm{v,i}\}}=D_{\{\mathrm{v,i}\}}^0\re^{-E_{\mathrm{m}{\{\mathrm{v,i}\}}}/T}$, where $E_{\mathrm{m}{\{\mathrm{v,i}\}}}$ is
the migration energy for vacancies and interstitials, respectively, $T$ is the
temperature. The last terms govern the nonlinear contribution caused by point
defects annihilation with the recombination coefficient $\alpha=4\pi
r_0(D_{\mathrm{i}}+D_{\mathrm{v}})/\Omega$ given in terms of recombination radius $r_0$ and atomic
volume $\Omega$.

In metallic systems due to the difference between migration energies of point
I- and V-defects (for example, for pure nickel $E_{\mathrm{mv}}=1.04$~eV,
$E_{\mathrm{mi}}=0.3$~eV with $r_0=1.5\cdot 10^{-9}$~m) there is a difference between
their diffusivities. It allows one to introduce a small parameter $\nu\equiv
D_{\mathrm{v}}/D_{\mathrm{i}}\ll 1$. In such a case, a renormalization of time scales allows us to
put $\nu\partial_{\mathrm{t}}c_{\mathrm{i}}\simeq0$ and eliminate the adiabatically fast variable $c_{\mathrm{i}}$
from the reduced second equation of the system (\ref{cvci}). Next, rewriting
$S_{\mathrm{v,i}}=Z_{{\{\mathrm{v,i}\}}N}\rho_{N}(1+\rho_{\mathrm{v}}^{*}+\rho_{\mathrm{i}}^{*})$, with
$\rho_{\mathrm{v,i}}^{*}\equiv\rho_{\mathrm{v,i}}/\rho_{N}$, and using dimensionless quantities
$t'\equiv t \lambda_{\mathrm{v}}$, $\lambda_{\mathrm{v}}\equiv D_{\mathrm{v}}Z_{\mathrm{v} N}\rho_{N}$, and
definitions $x_{{\{\mathrm{i,v}\}}}=\gamma c_{{\{\mathrm{i,v}\}}}$, $\gamma\equiv
\alpha/\lambda_{\mathrm{v}}$, $\mu\equiv(1+\rho^{*}_{\mathrm{v}}+\rho^{*}_{\mathrm{i}})$, $Z_{\mathrm{i} N}/Z_{\mathrm{v}
N}=1+B$, one can introduce $K'\equiv \gamma K/\lambda_{\mathrm{v}}$ measured in units of
displacement per atom. Then we drop primes for simplicity. In such a case, the
system (\ref{cvci}) is reduced to $\partial_{\mathrm{t}} x= K-\mu (x-x_0)- {K \nu
x}/[{\mu(1+B)}+ {\nu}x]$, where $x$ relates to the vacancy concentration. Here,
the last term is related to nonlinearity caused by interaction of interstitials
and vacancies governing the corresponding annihilation processes. In our
previous study (see reference \cite{EPJB2012}) we neglected the effect of interstitials
assuming their fast dynamics and motion to sinks. An allowance for
interstitials in the following derivation of the model does not change the
qualitative picture of the microstructure evolution, but this allowance
encumbers the physical model. This statement was discussed previously in
reference~\cite{Ryazanov}.

We have to note that the obtained one-component model does not incorporate
the production of defects by an elastic field caused by the presence of defects. Following
reference \cite{JETP96} this effect can be considered by introducing an additional
source term into dynamical equation for $x$ as $\re^{-E/T}$, where
$E=E_{\mathrm{f}}-\phi(\overline{r})$. Here, $\phi(\overline{r})$ is the potential of the
deformation field caused by the presence of defects. It facilitates the processes of defect
generation by decreasing the activation energy $E$; $E_{\mathrm{f}}$ is the defect
formation energy in the absence of deformation field;
$\overline{r}=x^{-1/3}$ is the averaged distance between the defects. At
$\overline{r}\to 0$, the field $\phi$ has the asymptotics \cite{JETP96}:
$\phi\simeq E^\re_0\;\overline{r}^3/(1+\overline{r}^6)$. Therefore, the
source term of defects can be rewritten as follows: $G\exp[\varepsilon x/(1+x^2)]$, where
$\varepsilon\equiv 2ZE^\re_0/T$ is defined through the defect formation energy
$E^\re_0\simeq 0.01$~eV and coordination number $Z$; the renormalized constant
$G$ is proportional to the probability of defect generation by an elastic field
caused by the presence of defects. It is determined by means of Debye frequency and atomic
volume; it exponentially depends on the relation between the defect migration energy
and temperature. The introduction of an additional term means that the deformation
field created by defects facilitates the defect generation due to a decrease in the
activation energy. This is essential in laser radiation due to temperature
instabilities, whereas at particle irradiation its efficiency is small
compared to the defect production in cascades. However, in our case, without loss
of generality, we retain this term assuming $G\ll 1$.

As far as point defects are mobile species of a microstructure the
corresponding rate equation would include spatial operators defined by
introducing a flux of defects. This diffusion flux has an ordinary Fick component
$-L_{\mathrm{d}}^2\pmb\nabla x$ with diffusion length $L_{\mathrm{d}}^2\equiv D_{\mathrm{v}}/\lambda_{\mathrm{v}}$ and
component describing the motion of defects with the velocity
$\mathbf{v}=(L_{\mathrm{d}}^2/T)\mathbf{F}$, $\mathbf{F}=-\pmb\nabla U$, where $U$ describes
the interaction of defects. In such a case, for the total flux, one has
\begin{equation}\label{J}
\mathbf{J}=-L_{\mathrm{d}}^2\pmb\nabla x+\mathbf{v}x.
\end{equation}
It can be rewritten in the canonical form $\mathbf{J}=-L_{\mathrm{d}}^2M(x)\pmb\nabla\mu(x)$,
where $M(x)=x$ is the mobility; $\mu(x)=\delta\mathcal{F}/\delta x$ plays the
role of chemical potential, where the free energy functional
\begin{equation}\mathcal{F}=\int{\rm d}\mathbf{r}f[x(\mathbf{r})]-\frac{1}{2T}\int{\rm
d}\mathbf{r}\int{\rm d}\mathbf{r}'x(\mathbf{r})\tilde
u(\mathbf{r},\mathbf{r}')x(\mathbf{r}')
\end{equation}
has the density $f(x)=x(\ln x-1)$. Here, the second part relates to pair
interactions in a self-consistent manner: $U(r)=-\int u(r-r')x(r'){\rm d}r'$
\cite{BHKM97,PhysicaD,EPJB2012}. We assume the attraction potential $-\tilde u(r)$ in a symmetric form, i.e., $\int \tilde u(r) r^{2n+1}{\rm d}r=0$. If the
field $x(r)$ does not vary essentially on the interaction range of defects
$r_0\simeq \Omega^{1/3}$, then one can use the approximation
\begin{equation}\label{expansionU}
\int{\rm d}\mathbf{r}' \tilde u(\mathbf{r}-\mathbf{r}')x(\mathbf{r}')\simeq
\varepsilon (x+ r_0^2\pmb\nabla^2x).
\end{equation}
The first term in equation (\ref{expansionU}) leads to the standard relation between
$U$ and $x$ in the framework of the elasticity theory \cite{UFN1996}. Indeed,
the effective flux takes the form $\mathbf{J}=-D(1-\varpi^2\kappa x/T)\pmb\nabla
x$, where $\kappa$ is the bulk modulus, $\varpi$ is the dilatation parameter.
The second term in the above expansion is responsible for microscopic properties of
defect interactions described by interaction radius $r_0$. Under normal
conditions, this term is negligible compared to the ordinary diffusion one.
However, as far as one has a density dependent diffusion coefficient
[$D(1-\varpi^2\kappa x/T)$], it can be negative in some interval for $x$. It
means that a homogeneous distribution of defects, starting with some critical
speed of its formation related to the temperature, sinks the density, and
the dilatation volume becomes unstable. The emergence of a directional flux of
defects results in supersaturation of vacancies and in the formation of voids. From
mathematical viewpoint such a divergence appearing at short time scales cannot be
compensated by a nonlinear part of the dynamical equation for $x$. The second
term in the expansion (\ref{expansionU}) can prevent divergencies of the derived
model due to microscopic properties of defect interactions. Therefore, the term
with the second derivative should be retained. This term governs the typical
sizes of defect clusters.

Inserting the above expressions into a rate equation for a vacancy concentration we
arrive at a deterministic reaction-diffusion equation of the form
\begin{equation}\label{le16}
\partial_{\mathrm{t}}x=R(x)- \pmb\nabla\cdot\mathbf{J}
\end{equation}
with
\begin{align}
%\begin{split}
R(x)&\equiv K-\mu (x-x_0)- \frac{K\nu x}{\mu(1+B)+ \nu x}+G
\exp\left({\frac{\varepsilon x}{1+x^{2}}}\right),\nonumber\\
\mathbf{J}&\equiv-\left[
 \pmb\nabla x -\varepsilon x\pmb\nabla(x+\ell^2\nabla^2x ) \right].
%\end{split}
\end{align}
Here, we use renormalization of a spatial coordinate as
$\mathbf{r}'=\mathbf{r}/L_{\mathrm{d}}$ and introduce a dimensionless length $\ell=r_0/L_{\mathrm{d}}$.

Considering real systems we have to take into account the effect of fluctuations
or stochastic sources representing the microscopic action onto the system
dynamics. In our further study we introduce a random source treated as an
internal noise resulting in dissipative dynamics of the whole system. Such
a noise should satisfy the fluctuation dissipation relation. Acting in the
standard manner, we rewrite the original deterministic model in the form
\begin{equation}\label{OCmodel1}
\partial_{\mathrm{t}} x=R(x) + \pmb\nabla\cdot M(x)\pmb\nabla\mu\,,
\qquad
\mu\equiv\frac{\delta \mathcal{F}}{\delta x}\,,
\end{equation}
where for the free energy functional $\mathcal{F}[x]$ we have
\begin{equation}
 \mathcal{F}[x]=\int{\rm d}{\mathbf{r}}\left[x\ln
x-x-\frac{\varepsilon}{2}x^2+\frac{\varepsilon\ell^2}{2}(\pmb\nabla x)^2\right].
\end{equation}
Formally, equation (\ref{OCmodel1}) can be represented in the canonical form
\begin{equation}\label{OCmodel_2}
 \partial_{\mathrm{t}} x=-\frac{1}{M(x)}\frac{\delta \mathcal{U}}{\delta x}\,,
\end{equation}
where for the functional $\mathcal{U}[x]$ we know only its first derivative,
i.e.,
\begin{equation}
\delta \mathcal{U}=-\int{\rm d}\mathbf{r}\delta x
\left\{M(x)R(x)+M(x)\pmb\nabla\cdot\left[M(x)\pmb\nabla\mu\right]\right\}.
\end{equation}
Following references \cite{PhysicaD,Haeften2004,MHW2005} we can introduce a fluctuation
source obeying the fluctuation-dissipation relation in an \emph{ad hoc} form:
\begin{equation}\label{OCmodel2}
 \partial_{\mathrm{t}} x=-\frac{1}{M(x)}\frac{\delta \mathcal{U}}{\delta x}+\sqrt{\frac{1}{M(x)}}\xi(\mathbf{r},t),
\end{equation}
where $\xi$ is white noise with $\langle\xi(\mathbf{r},t)\rangle=0$, $
\langle\xi(\mathbf{r},t)\xi(\mathbf{r}',t')\rangle=2\sigma^2\delta(\mathbf{r}-\mathbf{r}')\delta(t-t')$;
$\sigma^2$ is the noise intensity proportional to the bath temperature. Hence,
equation (\ref{OCmodel2}) treated in the Stratonovich interpretation represents the
generalized model considered below.

\section{Stationary states analysis}

In our study, the main attention is paid to stationary
patterns. Therefore, to describe the stochastic system behavior in the
stationary limit we have to obtain a stationary distribution
$\mathcal{P}_{\mathrm{s}}[x]$. To this end, we need to find a stationary solution of the
corresponding Fokker-Planck equation satisfying the Langevin equation
(\ref{OCmodel2}) \cite{Risken}. As it was shown previously (see
references \cite{MHW2005,PhysicaD,GO2001,WBL06,PhysA2008,EPJ2008,PhysA2009}) the
functional of the stationary distribution of the vacancy concentration field
has the explicit form
\begin{equation}
\mathcal{P}_{\mathrm{s}}[x]\propto \exp(-\mathcal{U}_\textrm{ef}[x]/\sigma^2)\, ,
\end{equation}
where the effective potential is
\begin{equation}\label{Ueff}
\mathcal{U}_\textrm{ef}[x]=\mathcal{U}[x]-\Sigma\int{\rm d}\mathbf{r}\ln M(x)\, .
\end{equation}
Here, $\Sigma$ is a renormalized parameter proportional to $\sigma^2$. It is
seen that the second term in the effective potential (\ref{Ueff}) serves as
entropy contribution. It may lead to the so-called \emph{entropy-driven} phase
transitions \cite{GO2001,EPJ2008,PhysA2009,PRE2004}, phase decomposition
\cite{PhysA2008,EPJB2010} and patterning \cite{MHW2005,WBL06,PhysicaD}.

\begin{figure}[!t]
\centerline{
\includegraphics[width=0.45\textwidth]{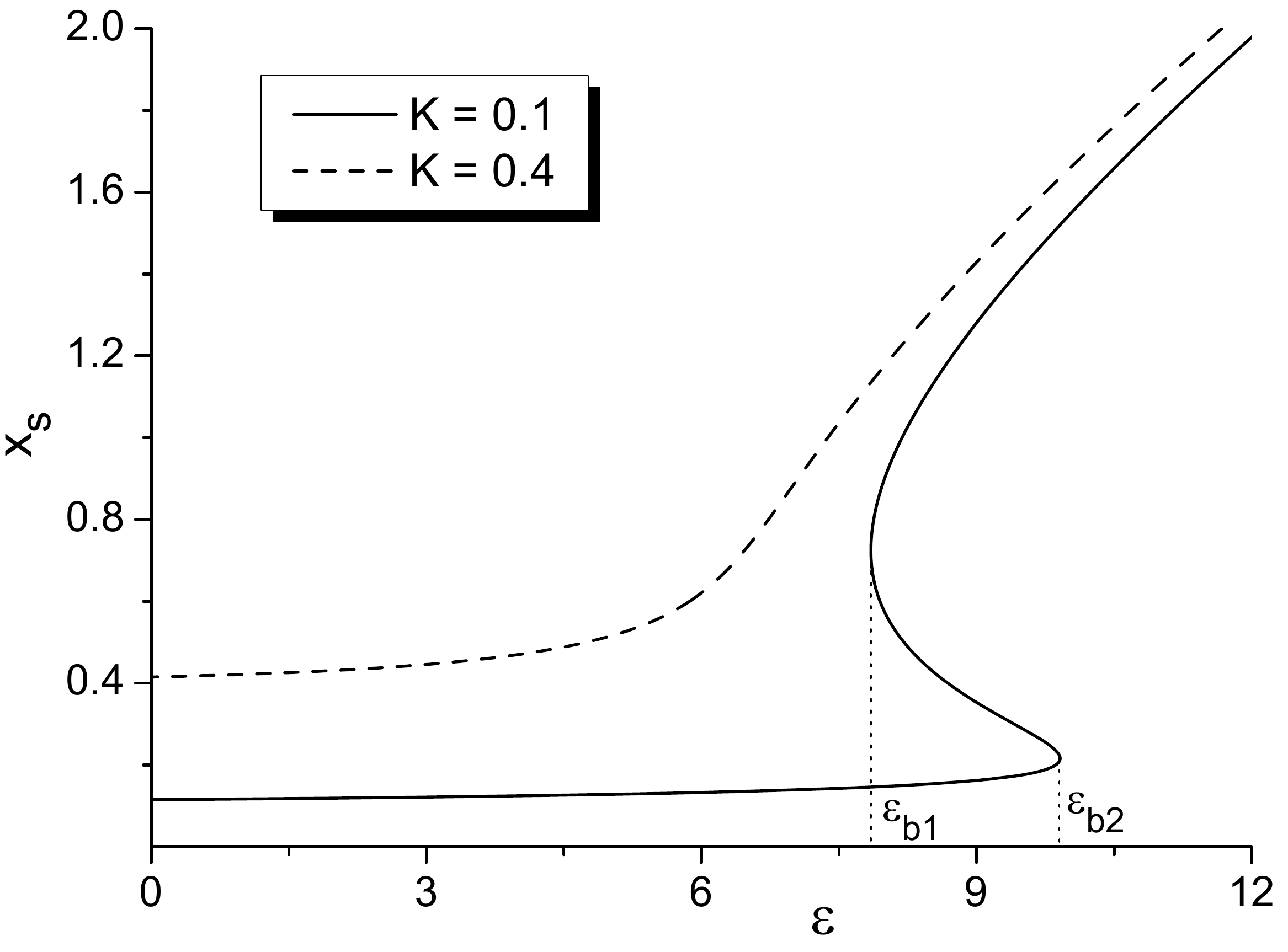}
\hspace{5mm}
\includegraphics[width=0.45\textwidth]{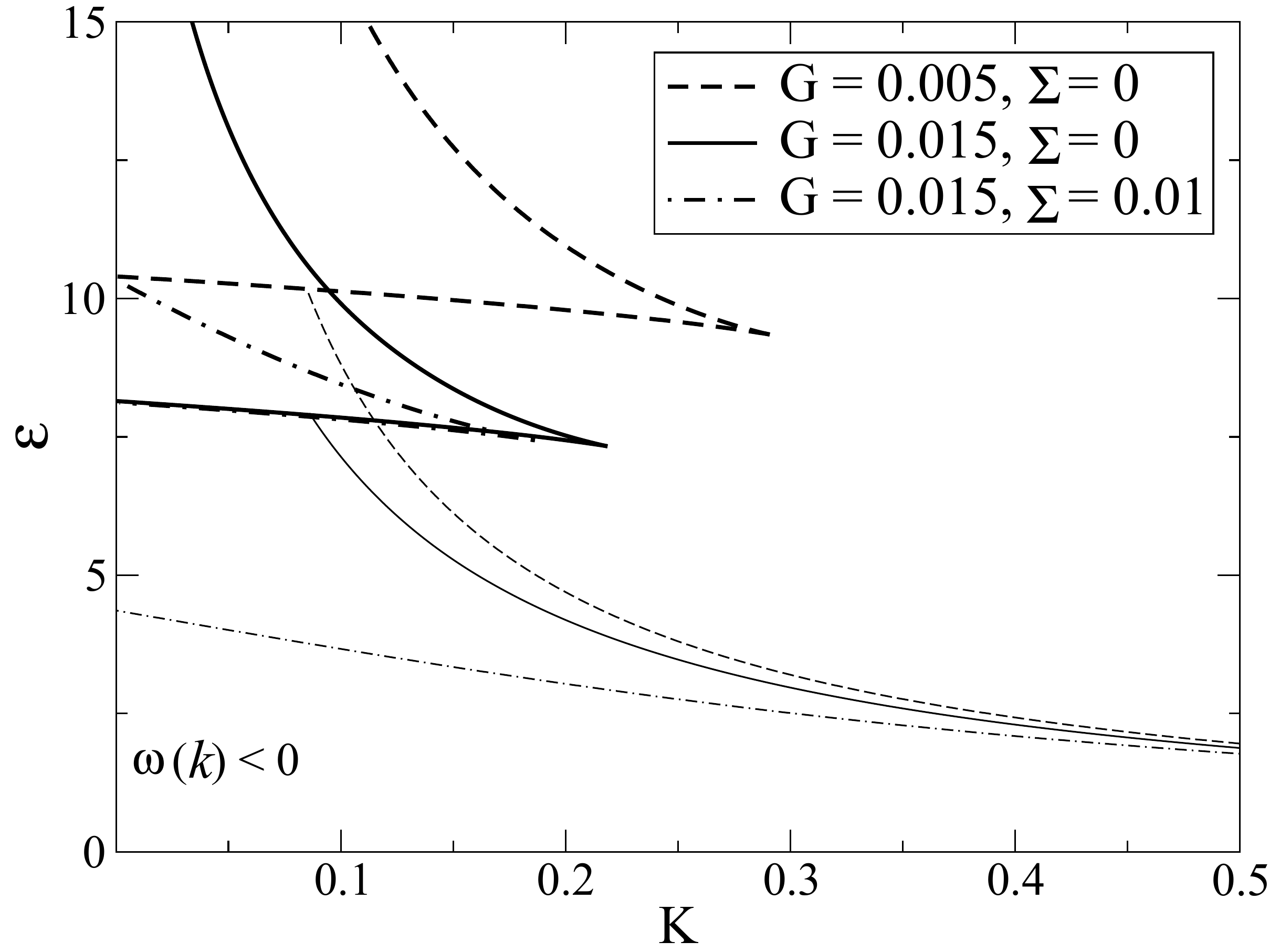}
}
\centerline{
(a) \hspace{0.475\textwidth} (b)
}
\caption{((a) Stationary dependencies of the point defects population at different values for $K$ at $G=0.015$, $\Sigma=0$. (b) Phase diagram at
different values for $G$ and $\Sigma$. Thick lines are spinodals. The corresponding
thin lines relate to $\omega(k)=0$ obtained for different $G$ and $\Sigma$.
\label{phd_0}}
\end{figure}

Firstly, let us consider the stationary states $x_{\mathrm{s}}$ of the homogeneous system
determined as solutions of the equation
\begin{equation}\label{stat_x}
R(x)+\frac{\Sigma}{M^2(x)}\frac{{\rm d }M(x)}{{\rm d}x}=0.
\end{equation}
The corresponding solutions are shown in figure~\ref{phd_0}~(a). It is seen that at
large defect damage rate the system is always in monostable state, whereas at
small $K$ the bistable regime is observed. The last effect means that at small
$K$  the formation of defects by elastic field plays an essential role and leads to
bistability of a system. At large $K$, the main mechanism of defect
production relates to defect generation in cascades. At small $K$, the
bistability domain for $\varepsilon$ at fixed $K$ is
$[\varepsilon_{b1},\varepsilon_{b2}]$. In the plane $(K,\varepsilon)$ binodals
$\varepsilon_{b1}(K)$ and $\varepsilon_{b2}(K)$ form a cusp binding the
bistability of the system states [see figure~\ref{phd_0}~(b)]. Here, in the cusp, the
system is bistable, while outside the cusp the system is monostable. Below the cusp,
the system is considered to be in a depleted state of defects, while above the cusp, an enriched state of defects is realized. For a homogeneous system, the depleted
state can be related to a ``crystalline'' phase with a small amount of defects,
whereas enriched state corresponds to some kind of ``amorphous'' phase with
a large number of defects. Therefore, in the cusp, two possible phases
(``crystalline'' and ``amorphous'') can be observed\footnote{To distinguish
real crystalline and amorphous states the long range order parameter should be
used. In our approach we do not have such a criteria. Therefore, two macroscopic
phases can be distinguished only by the value of the population
$x_{\mathrm{s}}$ of stationary defects.}.

As far as the functional of stationary distribution $\mathcal{P}_{\mathrm{s}}[x]$ is
known, the solution of the variational problem $\delta
\mathcal{P}_{\mathrm{s}}[x]/\delta x=0$ allows one to find stationary structures
$x_{\mathrm{s}}(\mathbf{r})$ as its solution. Following
references \cite{Haeften2004,MHW2005,PhysicaD,Mangioni2010} the most probable
solutions $x(\mathbf{r})$ corresponding to the minima of $\mathcal{U}_\textrm{ef}[x]$ can
be found as solutions of the equation
\begin{equation}\label{mps}
\partial_{\mathrm{t}}  x =-\frac{1}{M(x)}\frac{\delta \mathcal{U}_\textrm{ef}[x]}{\delta x}\,.
\end{equation}
Substituting the necessary expressions we get
\begin{equation}\label{mps1}
\partial_{\mathrm{t}}  x = R(x) + \pmb\nabla\cdot
M(x)\pmb\nabla\mu+\frac{\Sigma}{M^2(x)}\frac{{\rm d }M(x)}{{\rm d}x}\,.
\end{equation}

According to this equation and stationary states behavior we can study
the stability of stationary states in linear analysis by introducing small
fluctuations $\delta x=x-x_{\mathrm{s}}$. In such a case, the linearized equation
(\ref{mps1}) has an exponential solution $\delta x\propto \re^{[\Lambda
+\omega(k)]t}$, where $\Lambda$ is the Lyapunov exponent responsible for
homogenous perturbations, whereas $\omega(k)$ takes care of the stability of
stationary states due to inhomogeneous perturbations. It gives dispersion law
allowing one to define critical wave-numbers $k_\textrm{c}$ as solutions of equation
$\omega(k)=0$. Expressions for $\Lambda$ and $\omega(k)$ are:
%\begin{equation}
\begin{align}
\Lambda(x_{\mathrm{s}})&=-\mu-{\frac {K\nu\,\mu\,\left (1+B \right )}{\left
(\mu+\mu\,B+x_{\mathrm{s}}\nu\right )^{2}}}
+G\frac{\varepsilon(1-x_{\mathrm{s}}^2)}{(1+x_{\mathrm{s}}^2)^2}\exp\left(\frac{\varepsilon x_{\mathrm{s}}
}{1+x_{\mathrm{s}}^2}\right)-\frac{2\Sigma}{x^3_{\mathrm{s}}}\,,\nonumber \\
\omega(k;x_{\mathrm{s}})&=-k^2\left[1-\varepsilon
x_{\mathrm{s}}\left(1-\ell^2k^2\right)\right].
\end{align}
%\end{equation}
It follows that the noise action stabilizes the stationary state with respect
to homogeneous perturbations. In figure~\ref{phd_0} thin lines bind a domain of
inhomogeneous distribution of vacancies: at $\omega(k)<0$ no patterns can be
realized (below thin curves), in the opposite case dissipative structures of
point defects emerge (above thin curves). The analysis of the dispersion relation
with respect to three different solutions of equation (\ref{stat_x}) realized in
bistable domain allows one to set that for stable solutions $x_{\mathrm{s}}$, the noise
extends the domain of unstable modes whereas in the vicinity of  unstable branch
$x_{\mathrm{s}}$, the noise shrinks the domain for wave-numbers related to unstable modes.
Moreover, here the noise is capable of decreasing the value for the most unstable mode
$k_0=k_\textrm{c}/\sqrt{2}$.

\section{Numerical results}

Let us study the spatial arrangement of defect structure by means of
computer simulations. To this end, we make discretization of the system in
3-dimensional space with $N\times N\times N$ sites, where $N=128$. Using data
for pure nickel with the network dislocation density $\rho_N=10^{14}$~cm$^{-2}$
and diffusion length $L_{\mathrm{d}}\simeq 10^{-7}$~m as a reference system we can define
the total length of the system $L=N\Delta l$, with the mesh size $\Delta
l=0.5$. It provides the linear size of our 3D-system $L=25L_{\mathrm{d}}$. The time step
satisfying the stability of the simulation algorithm is $\Delta t=10^{-4}$ in
dimensionless units. Physically it corresponds to the time during which relaxation of one
cascade is finished, i.e., $10^{-8}$~s. Boundary conditions are periodic. To
study the dynamics of pattern formation we have  numerically solved the equation
(\ref{mps1}). At large time scales it gives the solutions equivalent in
statistical sense to solutions of the Langevin equation (\ref{OCmodel2}).

\begin{figure}[!b]
\centerline{
\includegraphics[width=0.9\textwidth]{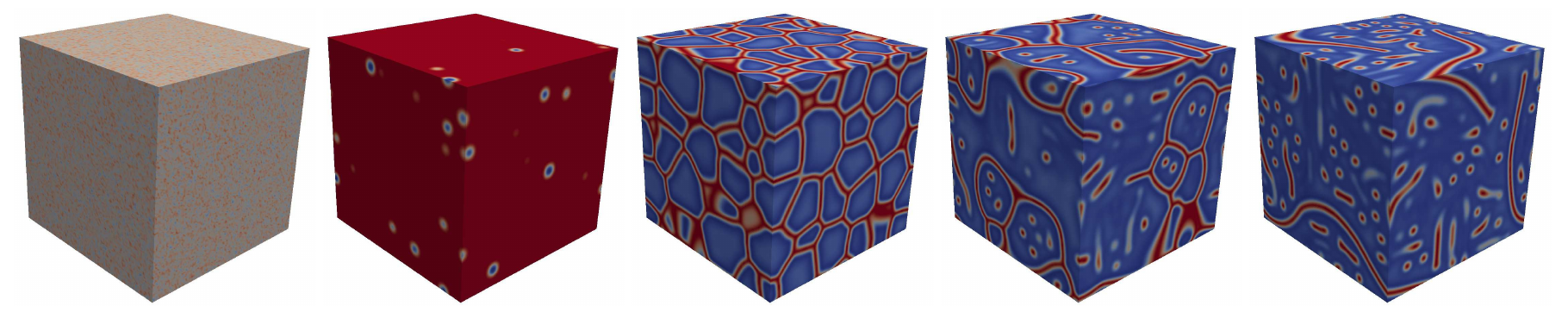}
}
\centerline{ (a) }
\centerline{
\includegraphics[width=0.9\textwidth]{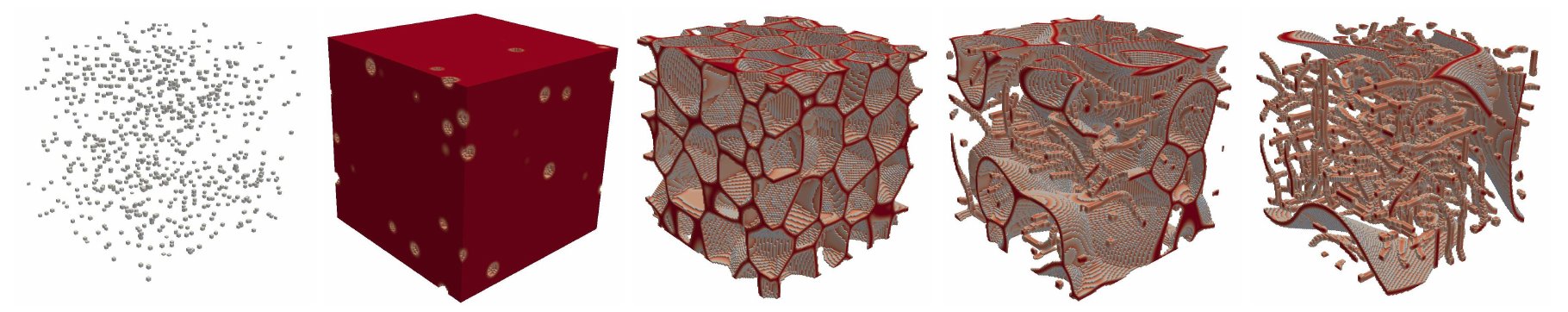}
}
\centerline{  (b) }
\caption{(Color online)
Snapshots of the system evolution obtained at $t=0$, 1, 10, 100, 300 (from left
to right). (a)
Total concentration field: blue domains relate to small concentration values, red domains correspond to high concentration of vacancies. (b) Spatial patterns of defects at the same times. Snapshots are taken at $K=0.075$, $\varepsilon=10.5$, $\Sigma=0$.
\label{evolK075}}
\end{figure}

In order to accelerate computations we exclude initial stages when vacancy
concentration increases from its equilibrium value by taking elevated initial
conditions $\langle x(\mathbf{r},t=0)\rangle=0.5$, $\langle(\delta
x)^2\rangle=0.1$. Simulation procedures were done using GRID infrastructure of
Ukrainian National Grid in the virtual organization MULTISCALE on computer
clusters \url{www.iap.sumy.org} (Institute of Applied Physics, NAS of
Ukraine) and \url{www.icmp.lviv.ua} (Institute for Condensed Matter Physics,
NAS of Ukraine).

Let us consider the behavior of a deterministic system, firstly, setting
$\Sigma=0$. Snapshots of the typical evolution of the system at $K=0.075$ are
shown in figure~\ref{evolK075}. Here, in figure~\ref{evolK075}~(a) the total distribution
of the vacancy field is shown [blue domains (online) relate to a small vacancy
concentration limit, red ones (online) correspond to high concentration limit].
In figure~\ref{evolK075}~(b) spatial defect structures are shown as a cross-section
at a fixed concentration value $x=0.9$.
Here, the concentration of initially Gaussianly distributed defects is small.
During the system evolution, the concentration increases essentially. It is
stipulated by the production of defects through irradiation and elastic field in the system
having an annealed concentration of defects at previous stages. When
supersaturation is reached, the defects start to condense into phases (structural
type of grain boundaries) with high concentration of defects. At next stages of
the system evolution one gets grains without defects. From thermodynamical
viewpoint such ``ideal'' grains are unstable and at the next time steps one can
observe the formation of additional defects inside the grains. Grains evolve
according to Ostwald ripening mechanism. At late stages corresponding to
stationary limit one has a net of vacancy loops and vacancy walls with voids.

\begin{figure}[!t]
\centerline{
\includegraphics[width=0.48\textwidth]{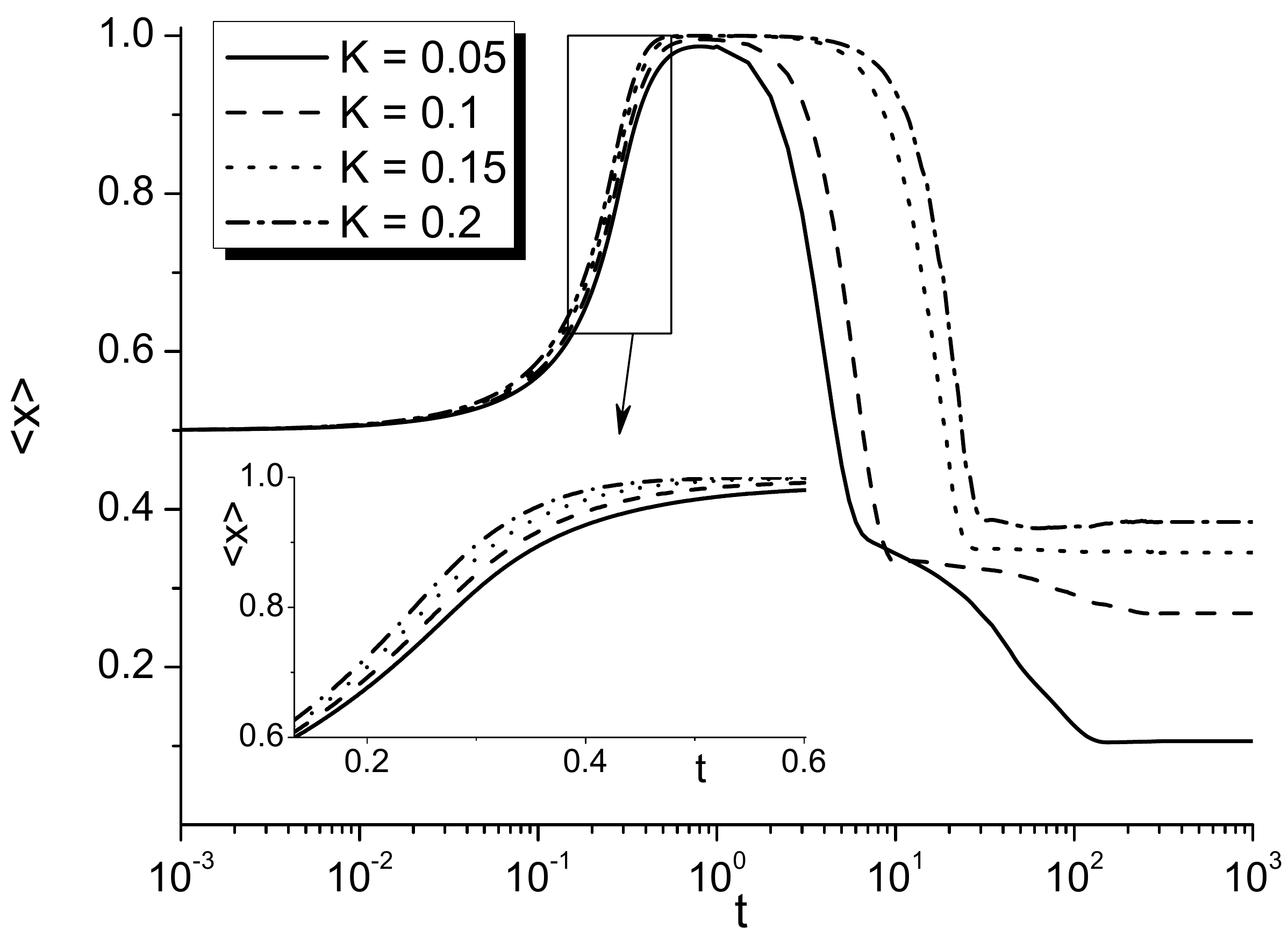}
\hspace{2mm}
\includegraphics[width=0.48\textwidth]{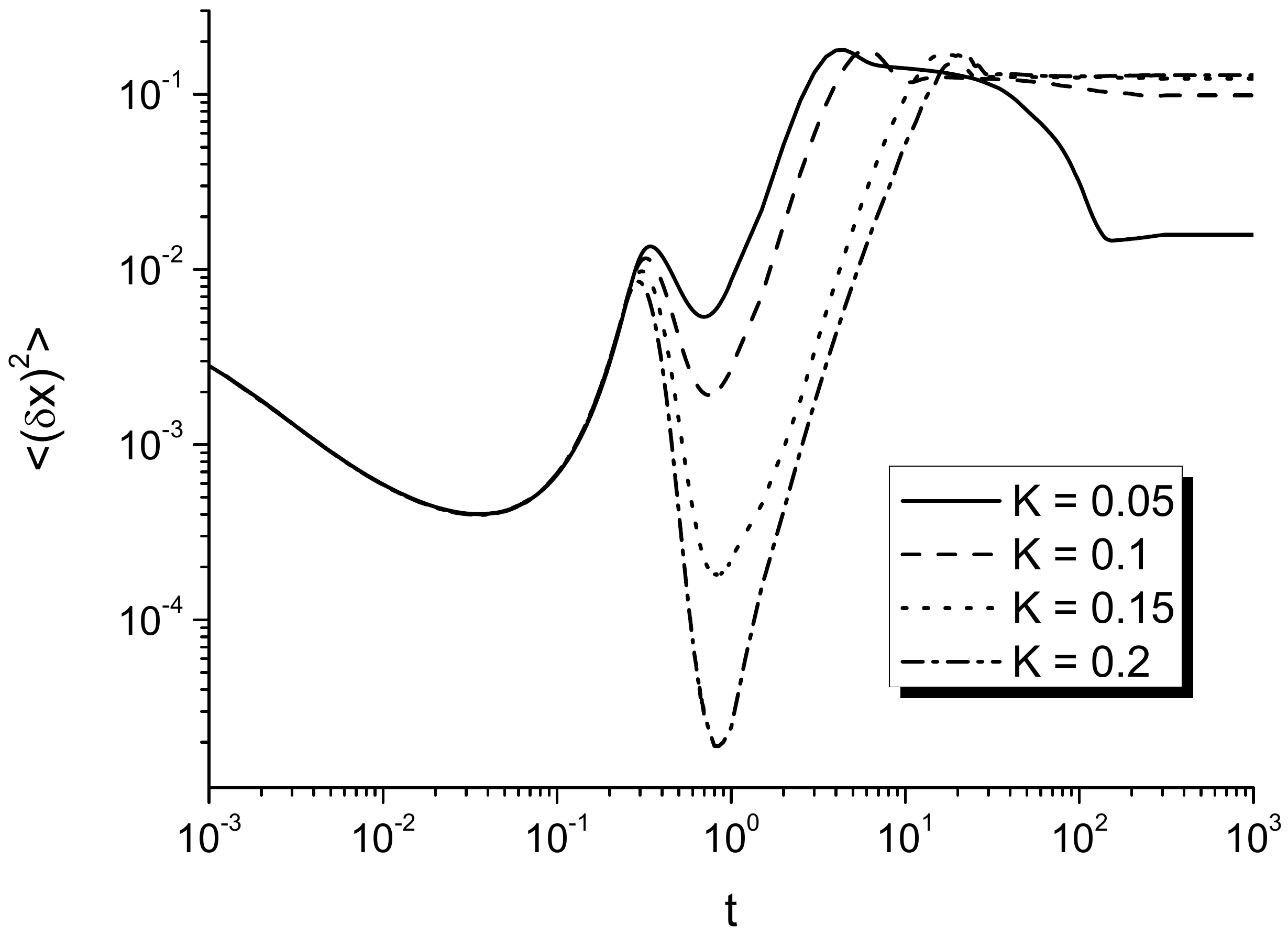}
}
\centerline{(a) \hspace{0.47\textwidth} (b)}
\vspace{2mm}
\centerline{
\includegraphics[width=0.7\textwidth]{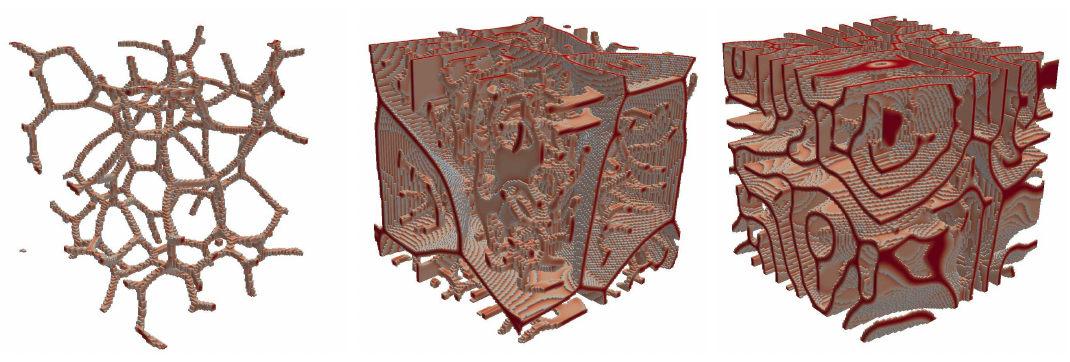}
}
\centerline{(c)}
\caption{(Color online) The dynamics of the
averaged concentration (a) and its dispersion (b) at different $K$ and
$\varepsilon=10.5$, and $\Sigma=0$. (c) Stationary defect structures at
$K=0.001$, 0.1, 0.2 (from left to right).\label{x(t)} }
\end{figure}

\begin{figure}[!t]
\centerline{
\includegraphics[width=0.5\textwidth]{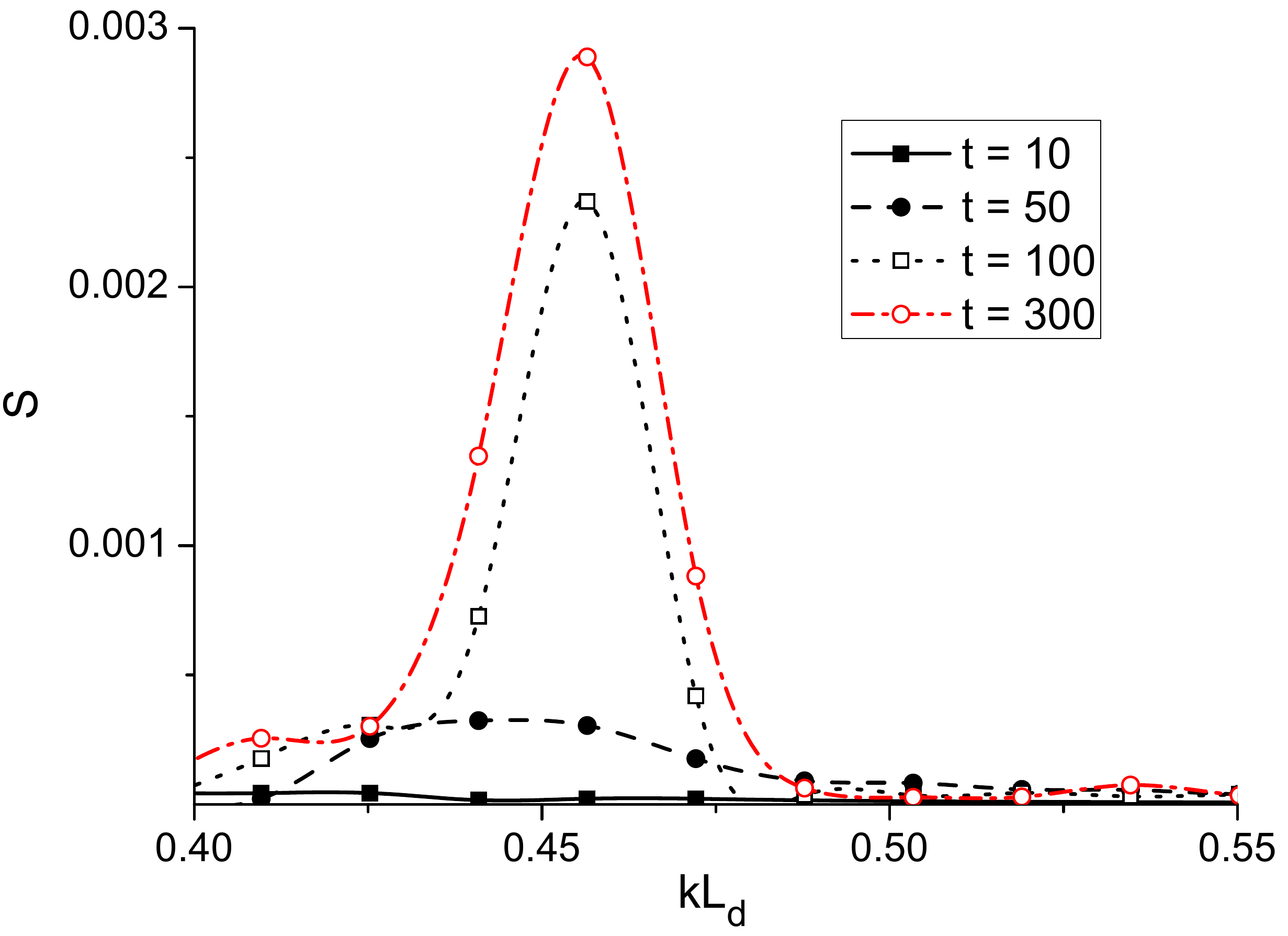}
}
\caption{(Color online) Spherically averaged structure function dynamics
at $K=0.1$, $\varepsilon=10.5$, $\Sigma=0$.
\label{S(k)}}
\end{figure}

\looseness=-1 The quantitative picture of the system evolution is shown in figure~\ref{x(t)}.
Here, the averaged concentration $\langle x\rangle$ [see figure~\ref{x(t)}~(a)]
increases at small times and after supersaturation it decreases toward
stationary value $\langle x\rangle_{\mathrm{s}}$ due to the formation of defect clusters and
motion of defects to sinks. It is seen that with the growth of the defect
production rate, an average concentration $\langle x \rangle$ takes up elevated
stationary values. The principal information about the ordering process is provided by
the dispersion behavior $\langle(\delta x)^2\rangle$ [see figure~\ref{x(t)}~(b)].
This quantity plays the role of an order parameter in phase transition, phase
separation and pattern formation processes
\cite{Garcia,PhysicaD,PhysA2010_1,CEJP2011}. An increasing dynamics of this
quantity means the ordering of the system with the formation of different phases (in
our case, phases with low and high vacancy concentrations). If it attains a
non-zero stationary value, then an ordered phase organizes. In our case it is
seen that $\langle(\delta x)^2\rangle$ increases toward maximal value
corresponding to the formation of well-structured grains: if grains have a small
number of defects inside, then the order parameter takes a large maximal value.
At late stages, due to reconstruction of defect clusters, it decreases toward its
stationary non-zero values. It is seen that with the growth of defect
production rate, the reconstruction of clusters is faster. Moreover, at
elevated $K$, the order parameter takes up larger values meaning the formation of well
organized structure of point defect clusters. The corresponding stationary
snapshots [see figure~\ref{x(t)}~(c)] illustrate different structures of defects at
different values for $K$.
The dynamics of spatial arrangement of defects can be seen from the structure
function $S(\mathbf{k},t)$ time behavior as the Fourier transformation of the
two-point correlation function $C(\mathbf{r},t)\equiv\langle \delta
x(\mathbf{r},t)\delta x(\mathbf{0},t)\rangle$. The spherical analogue of
$S(\mathbf{k})$ is shown in figure~\ref{S(k)}.
It is seen that during the system evolution, the peak of $S(k)$ that denotes the period
of spatial structures increases and shifts toward large wave-numbers $k$
meaning the formation of a structure with a smaller period compared to structures
related to the early stages when supersaturation is reached. An increase in
the peak height corresponds to the formation of well-organized phases enriched by
V-type defects.

\begin{figure}[!b]
\centerline{
\includegraphics[width=0.47\textwidth]{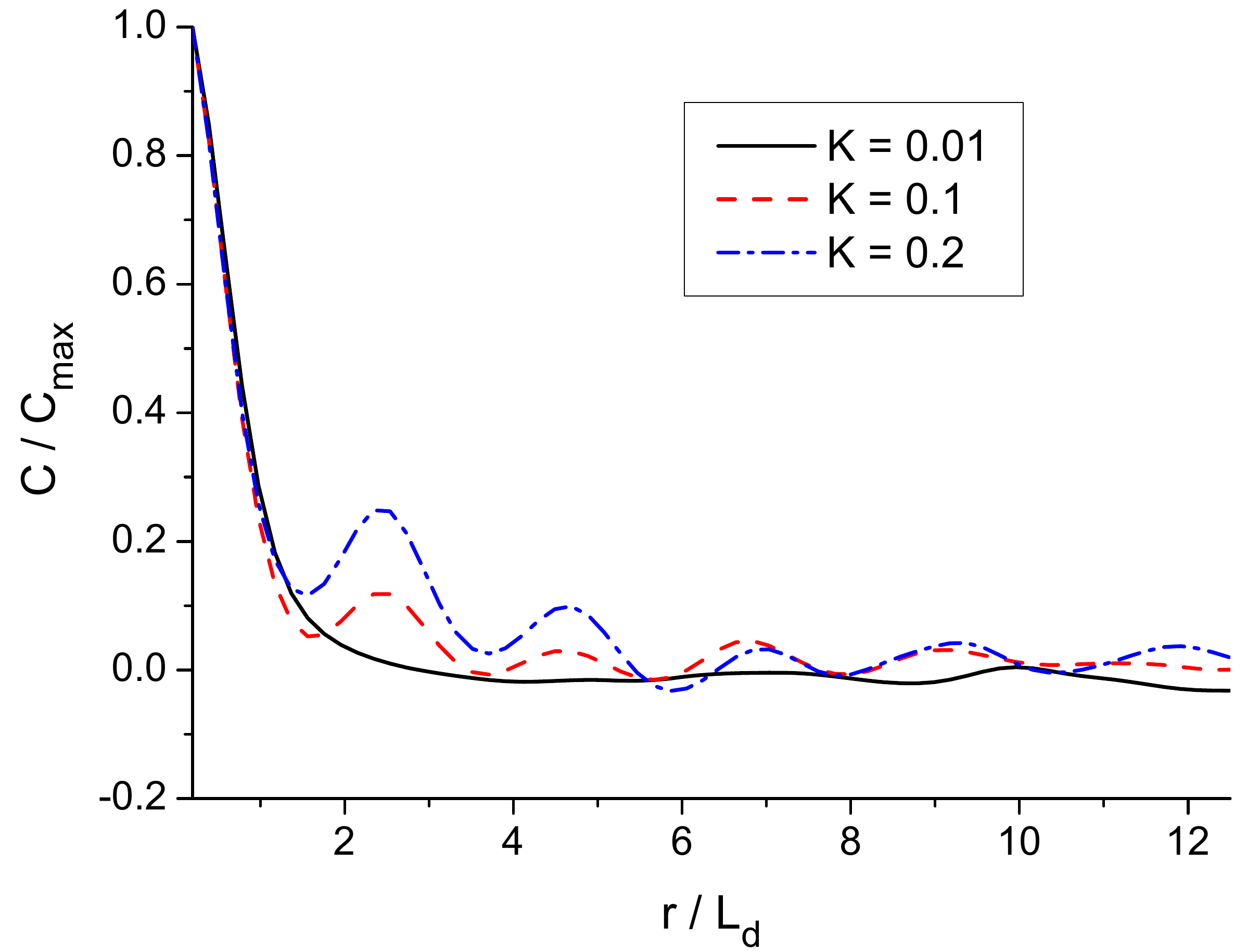}
\hspace{1mm}
\includegraphics[width=0.51\textwidth]{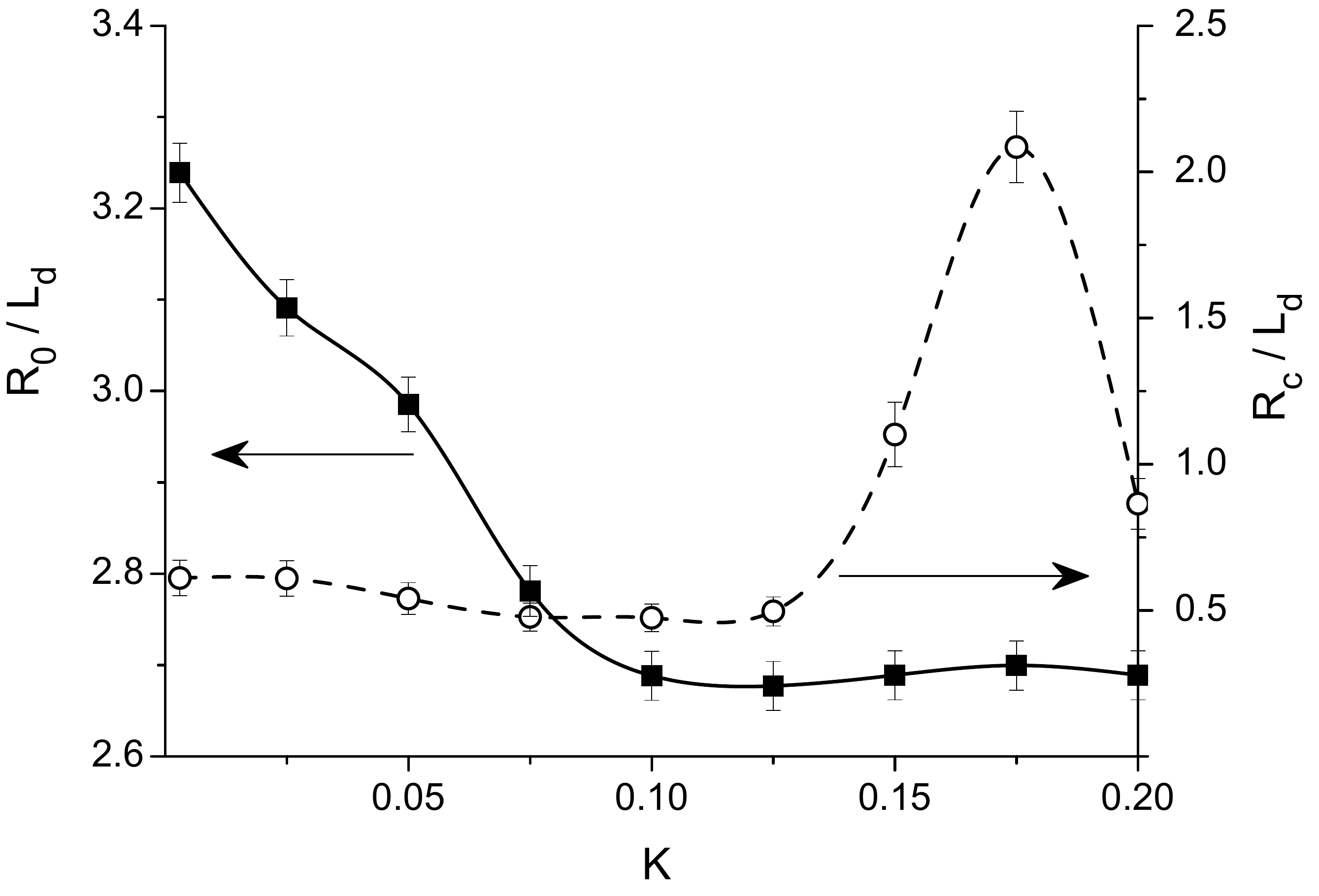}
}
\centerline{ \hspace{5mm}(a) \hspace{0.45\textwidth} (b) }
\vspace{2mm}
\centerline{
$K=0.025$\hspace{20mm}$K=0.075$\hspace{20mm}$K=0.175$
}
\centerline{
\includegraphics[width=0.7\textwidth]{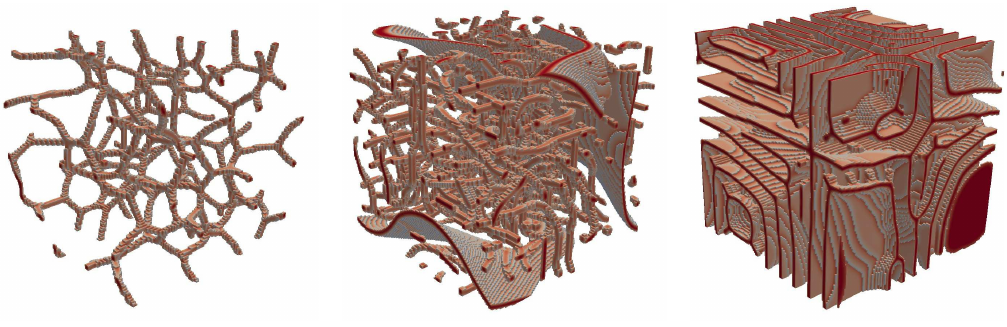}
}
\centerline{(c)}
\caption{(Color online) (a) Spherically averaged two-point correlation function $C(r)$
 for stationary patterns at $K=0.01$, 0.1, 0.2.
 (b) The dependence of correlation radius $r_\textrm{c}$  and period of patterns $r_0$
 \emph{vs} damage rate $K$. (c) Snapshots of stationary patterns illustrating microstructure at different values for
 their period $r_0$ and correlation radius $r_\textrm{c}$ at variation of the damage rate $K$.  Other parameters are: $\varepsilon=10.5$, $\Sigma=0$.
 \label{C(r)}}
\end{figure}

More information on spatial arrangement of V-type clusters in the
stationary limit is provided by a two-point correlation function $C(r,t\to\infty)$. It is
a spherical analogue calculated in 3D space at different values for defect
production rate shown in figure~\ref{C(r)}~(a).
It follows that $C(r)$ exponentially  decays from its maximal value at the
point $r=0$. The decaying rate depends on the damage rate $K$. It is
worth noting that with an increase in $K$, spatial correlation function
manifests an oscillatory behavior meaning the formation of ordered structures with
a fixed period. To study stationary patterns in detail we use an approximation
of the correlation function in the form $C(r,t\to\infty)\simeq
C_\textrm{max}\re^{-r/r_\textrm{c}}\cos(2\pi r/r_0)$, where $r_\textrm{c}$ is the correlation scale, $r_0$
relates to the period of the pattern. From the dependence $C(r)$ one can find
the correlation radius of defect structures indicating spatial interactions between
structural elements in the pattern. From the Fourier transform of $C(r)$ we can
obtain a period of patterns $r_0$. The corresponding dependencies $r_\textrm{c}$ and $r_0$
are shown in figure~\ref{C(r)}~(b). It follows that at small $K$ the correlation
function manifests decaying oscillations with small amplitude and large period,
where $C(r\gg1)\to 0$. In such a case we arrive at a network structure where all
V-defects are arranged into statistically independent loops [see left hand snapshot
in figure~\ref{C(r)}~(c)]. When we increase $K$, a well pronounced oscillatory behavior
of the correlation function with decaying correlations is realized. Compared
to the previous case, one can say that the period of spatial structures decreases,
and nonzero values for $C(r\gg1)$ indicate the formation of a pattern representing
defect walls with defect voids and loops [see centered snapshot in
figure~\ref{C(r)}(c)]. In the case of large $K$ $(K=0.175)$, the decaying rate for
$C(r)$ is characterized by large values for the correlation radius and small period
of structures. Such correlations are caused by the formation of planar structures
shown in figure~\ref{C(r)}~(c) (right hand snapshot). At a further increase in $K$, the
production of defects leads to the formation of defects inside planar structures resulting in the correlation radius decrease and a small decrease in the period of patterns
[see figure~\ref{x(t)}~(c) right hand snapshot]. The predicted behavior of the period of
patterns \emph{versus} damage rate is in good correspondence with well-known
experimental data \cite{Konobejev}.

\begin{figure}[!t]
\centerline{
\includegraphics[width=0.5\textwidth]{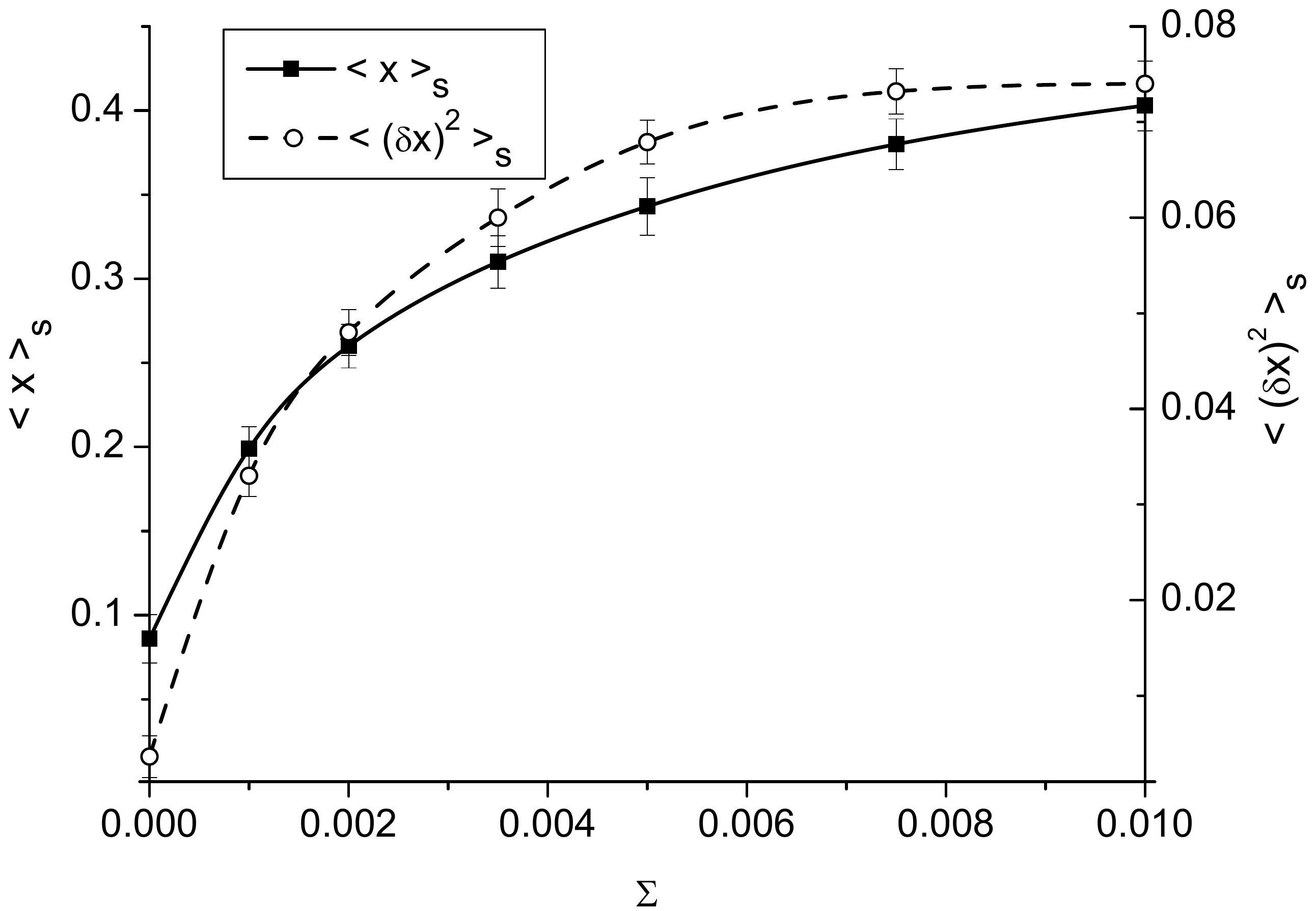}
}
\caption{Stationary
dependencies of the averaged vacancy concentration field $\langle x\rangle_{\mathrm{s}}$
and dispersion $\langle(\delta x)^2\rangle_{\mathrm{s}}$ on the noise intensity $\Sigma$.
Other parameters are: $\varepsilon=9$, $K=0.05$.\label{x_Sigma}}
\end{figure}

Next, let us study the most probable stationary patterns in stochastic case by varying
the noise intensity $\Sigma$. For a homogeneous system one can find most
probable values for the vacancy concentration as solutions to the stationary
homogeneous equation (\ref{stat_x}). It follows that the noise action leads to
an increase in the most probable vacancy concentration due to a stochastic
mechanism of defect production compared to the deterministic case. Studying
the stationary patterns by a numerical solution of the original equation (\ref{mps1}) one can
find dependencies of $\langle x\rangle_{\mathrm{s}}$ and $\langle(\delta x)^2\rangle_{\mathrm{s}}$
versus the noise intensity $\Sigma$. The corresponding results are shown in
figure~\ref{x_Sigma}.
It is seen that with the growth of noise intensities the averaged concentration
increases. The same effect is observed in the dependence $\langle(\delta x
)^2\rangle_{\mathrm{s}}(\Sigma)$. It means that large fluctuations in the  production
of defects promote the organization of well-ordered structures of V-type defects due to
an entropy-driven mechanism of pattern formation \cite{EPJB2012}.

\begin{figure}[!t]
\centerline{
\includegraphics[width=0.47\textwidth]{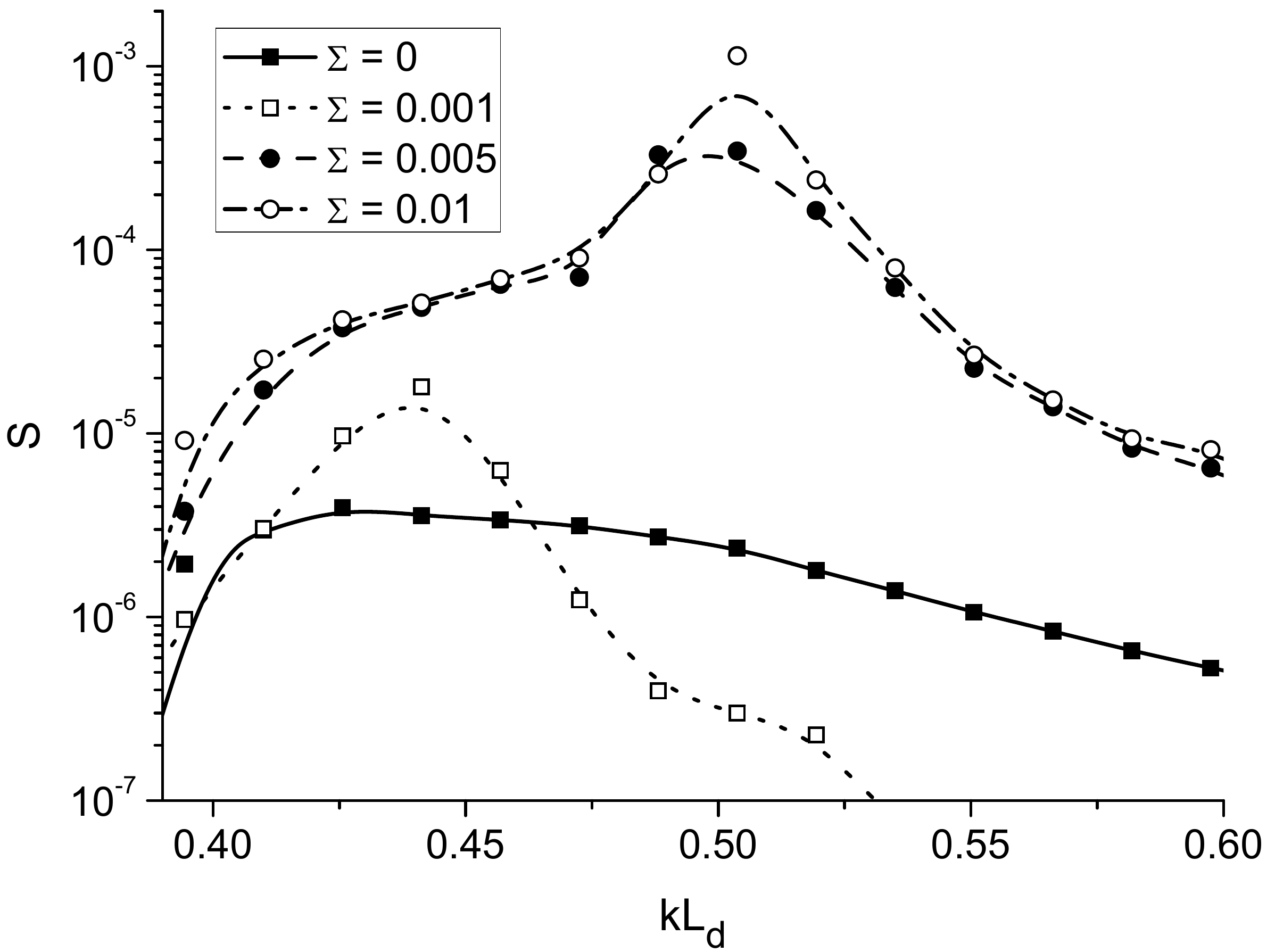}
\hspace{1mm}
\includegraphics[width=0.51\textwidth]{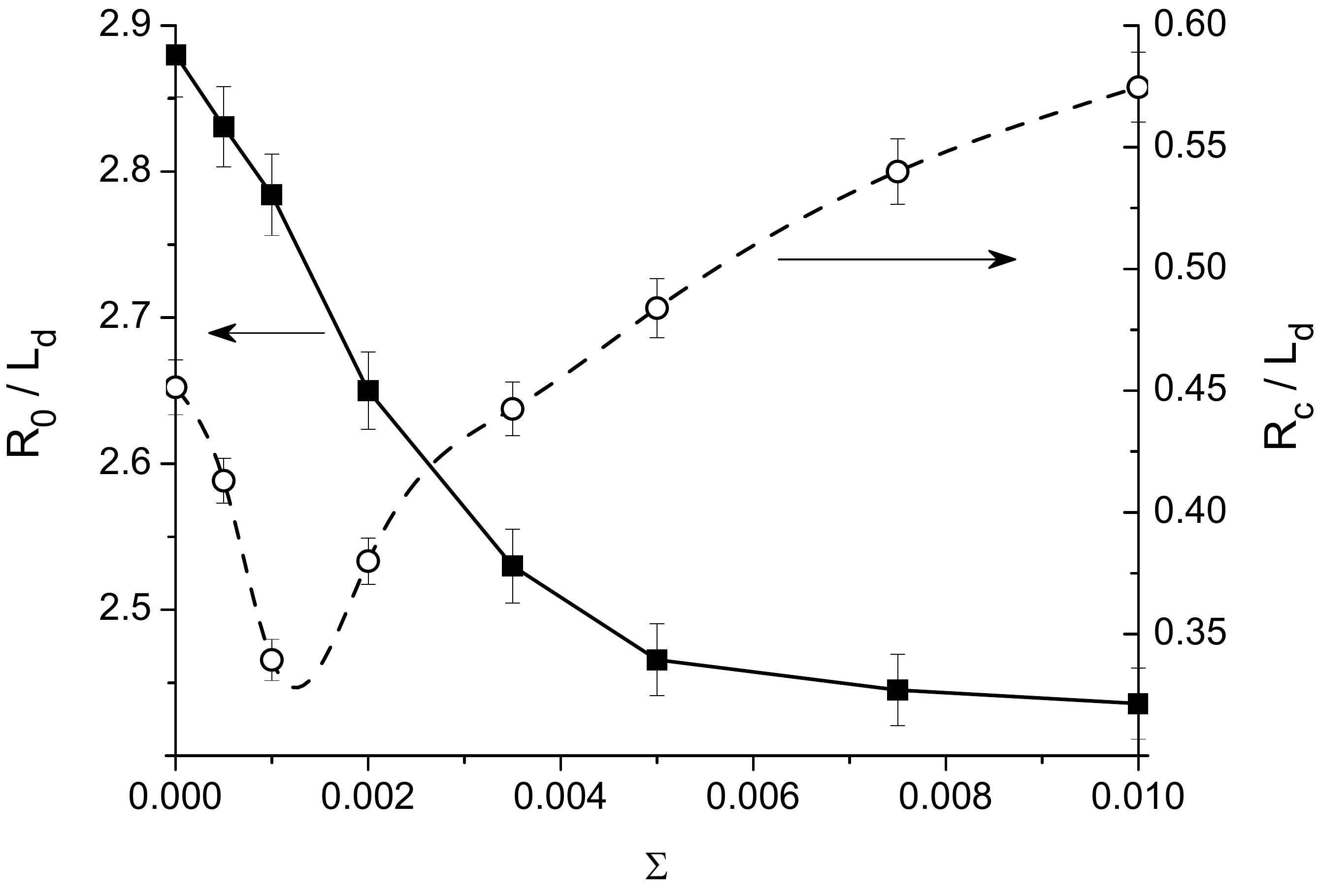}
}
\centerline{(a) \hspace{0.45\textwidth} (b)}
\vspace{2mm}
\centerline{ $\Sigma=0.0$\hspace{20mm} $\Sigma=0.001$\hspace{20mm} $\Sigma=0.005$}
\centerline{
\includegraphics[width=0.7\textwidth]{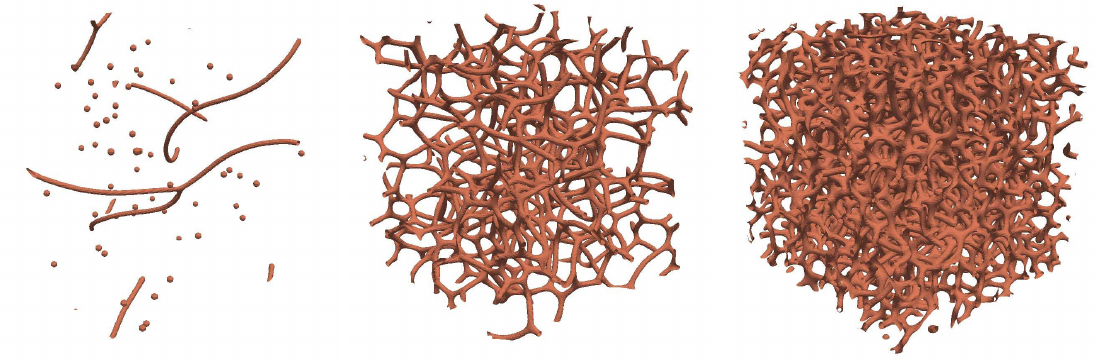}
}
\centerline{(c) }
\caption{(Color online) (a) Stationary
structure function at different noise intensity. (b) Dependencies of the
correlation radius $r_\textrm{c}$ and period of patterns $r_0$ \emph{vs} noise intensity
$\Sigma$. (c) Snapshots of stationary patterns obtained at $\Sigma=0.0$, 0.001
and 0.005 (from left to the right). Other parameters are: $\varepsilon=9$,
$K=0.05$. \label{s(k)Sigma}}
\end{figure}

From the naive consideration one can predict a decrease of the period of
structures and an increase in their correlation radius. Indeed, analyzing the
corresponding structure function shown in figure~\ref{s(k)Sigma}~(a) one can find
that with the noise intensity growth the position of its maximum is shifted
toward large wave-numbers, i.e., the period of defect structures decreases. A growth
of the peak at elevated $\Sigma$ corresponds to an increase in the correlation
radius. The corresponding dependencies of the correlation radius and the period of
patterns are shown in figure~\ref{s(k)Sigma}~(b). Snapshots of stationary patterns at
different noise intensities and fixed main system parameters are shown in
figure~\ref{s(k)Sigma}~(c) in order to show the morphology changes of patterns. It is seen
that in the noiseless case one has a large amount of voids and a small number of
linear defects. If we introduce the noise, then the effective potential
(\ref{Ueff}) determining the stationary distribution of the vacancy field
$\mathcal{P}_{\mathrm{s}}[x]$ acquires logarithmic component (i.e., entropy contribution), or
pre-exponential term in $\mathcal{P}_{\mathrm{s}}[x]$. Therefore, due to this effect one
gets a morphology change of patterns at an essential increase of the number of linear defects only if the noise term is introduced. In such a case, the emergence of a
net of linear defects results in a decrease of the correlation radius $r_\textrm{c}$ at
small $\Sigma$. With a further growth of $\Sigma$, one gets large correlations in
patterns in the directions of the developed linear defects.

\section{Discussion}

In our study we have considered the pattern formation in systems of point defects
under the action of irradiation. We have generalized a standard approach of point
defects dynamics by taking into account the production of defects by an elastic field
caused by the presence of defects, their interactions and stochastic
production of defects as a microscopic effect on the system described at
mesoscopic level.

The main result of our paper relates to identifying the effect of
irradiation conditions and microscopic processes of defect production and
their interactions onto morphology changes of the V-type defect structure.
Considering the deterministic picture of pattern formation in the framework of
correlation analysis we have shown that the period of patterns decreases with the
displacement damage rate growth. It is compared with the diffusion length and
according to the data for pure nickel with the net dislocation density
$\rho_N\sim10^{14}$~m$^{-2}$ it takes up values around $10^{-7}$~m. For a cold worked
nickel characterized by $\rho_N\sim10^{15}$~m$^{-2}$, the period of patterns is
around $100$~\AA. This result was discussed previously by experimental
investigations \cite{Konobejev}. We have studied the morphology changes of stationary patterns
by measuring the correlation radius of stationary structures. It was
shown that when the morphology of patterns changes crucially the correlation
radius manifests an anomalous behavior. In a deterministic system, this effect is
well seen when transition from linear structures toward the formation of planar
structures (defect walls or grain boundaries) is realized. Here, the correlation
radius takes up large values for pure planar defect structures.

Considering stationary patterns in a stochastic system it was shown that the
noise action plays a role similar to the displacement damage rate which increases
the number of defects. Due to supersaturation of point defects their
organization into clusters of higher dimension can be observed. We have
compared stationary patterns for a deterministic system and the corresponding
stationary patterns for a stochastic one. Here, the period of patterns
decreases with the noise intensity growth. We have shown that due to
reconstruction of stationary distribution of the vacancy concentration field
caused by the noise effect, the morphology of patterns changes
essentially: there is a transition from patterns with voids to patterns with
linear defects. Such a morphology change corresponds to a decrease in the
correlation radius of spatial structures. With further noise intensity growth,
spatial arrangement described by an increasing correlation radius is observed.

We have considered the dynamics of vacancies only assuming fast relaxation of
interstitials and neglecting the dynamics of other elements such as di-, tri-,
tetra-vacancies and dislocations. The proposed approach can be generalized by
taking into account the dynamics of all the above elements. In our study we have used
material constants for pure nickel. However, the obtained results, which are quite
general, can be applied to the study of an ensemble of point defects in certain material
under sustained irradiation.

\section*{Acknowledgement}

Support from National Academy of Sciences of Ukraine under grant on usage and
developing GRID-technology is gratefully acknowledged.

%\newpage
\ukrainianpart

\title{Особливості просторової організації дефектів вакансійного типу
 в опромінюваних системах: 3D--моделювання}%
\author{В.О. Харченко, Д.О. Харченко}
\address{Інститут прикладної фізики НАН України,  вул. Петропавлівська, 58, 40000 Суми, Україна}
\makeukrtitle
\begin{abstract}
\tolerance=3000%
Проведено дослідження динаміки структуроутворення в системі точкових дефектів
при сталій дії опромінення в рамках швидкісної теорії. Нами узагальнено
стандартний підхід врахуванням впливу пружних полів та стохастичного
виробництва дефектів, що представляється внутрішнім мультиплікативним шумом. В
рамках застосування процедури 3D-моделювання встановлено, що зростання
швидкості набору дози приводить до зміни морфології структур, які складаються з
вакансій. Аналогічний ефект спостерігається при варіюванні інтенсивності
мультиплікативного шуму. Стаціонарні структури дефектів досліджено із
застосуванням кореляційного аналізу.

\keywords дефекти, опромінення, структуроутворення, шум

\end{abstract}

\end{document}